\documentclass{svmult}
\usepackage[T1]{fontenc}
\usepackage[latin1]{inputenc}
\usepackage{mathptmx}
\usepackage{helvet}
\usepackage{courier}
\usepackage{makeidx}
\usepackage{multicol}
\usepackage{footmisc}
\usepackage{graphicx}
\usepackage{pgfplots}
\pgfplotsset{width=7cm,compat=1.8}

\begin{document}
\title*{The effect of a physical robot on vocabulary learning}
\author{Andreas Wedenborn \and Preben Wik \and Olov Engwall \and Jonas Beskow}
\institute{Jonas Beskow \at KTH Speech, Music and Hearing, \email{beskow@kth.se}}
\institute{Andreas Wedenborn \at KTH \and Preben Wik \at Furhat Robotics AB, \email{preben@furhatrobotics.com} \and Olov Engwall \at KTH Speech, Music and Hearing, \email{olov@kth.se} \and Jonas Beskow \at KTH Speech, Music and Hearing, \email{beskow@kth.se}}
\maketitle

\abstract{This study investigates the effect of a physical robot
  taking the role of a teacher or exercise partner in a language
  learning exercise. In order to investigate this, an application was
  developed enabling a 2:nd language learning vocabulary exercise in
  three different conditions. In the first condition the learner would
  receive tutoring from a disembodied voice, in the second condition
  the tutor would be embodied by an animated avatar on a computer
  screen, and in the final condition the tutor was a physical robotic
  head with a 3D animated face mask. A Russian language vocabulary
  exercise with 15 subjects was conducted. None of the subjects
  reported any Russian language skills prior to the exercises. Each
  subject were taught a set of 9 words in each of the three conditions
  during a practice phase, and were then asked to recall the words in
  a test phase. Results show that the recall of the words practiced
  with the physical robot were significantly higher than that of the
  words practiced with the avatar on the screen or with the disembodied voice.}

\section{Introduction}

Spoken multimodal interaction is rapidly becoming a key technology in
human robot interaction. One of the areas where social robots have
been expected to have large impact is in education. In a study by Scasselati
et al \cite{KEEPON}, children perfomed better at a puzzle-solving task when they
recieved tutoring from a physical robot than when they got identical
tutoring from an on-screen character or a disembodied voice.   

Computer aided language learning (CALL) \cite{CALL} is an area of
education where conversational agents have been employed with positive
results. Such systems can through the use of facial 
expressions, gestures, speech and dialogue skills enable a 
human-computer interaction that to some extent resembles face-to-face 
communication and as such provide language learners with virtual
teaching partners.  Wik and Hjalmarsson \cite{DEAL} have shown that the 
attitude towards embodied conversational agents as a complement to a
human teacher is positive.

In the present study, the goal is to investigate the effect of
physical embodiment in a simple language learning task, and wether the
effect seen in \cite{KEEPON} can be replicated in this setting.
 
\subsection{Virtual language teacher}

For this study, a simple spoken interactive language training exercise was
constructed. The aim of the exercise was to teach Russian
vocabulary to learners with no previous knowledge of
Russian.  Three versions of the exercise were constructed: one where
the learner is instructed by a disembodied voice, one with a talking
animated avatar on the screen, and one where the user was tutored by a
physical robot. The interaction language was English. 

In order to increase task motivation the exercise was constructed in a
social way where the user was not just presented with spoken playback of the word and then the picture and text, but
instead the task would be introduced through a narrative from the
teacher, and the user would be taken through the words
one by one, through varied dialogue, being given feedback along the
way. 

In each exercise, the learner was introduced to nine words. For each word they would hear the
pronunciation, see it in writing and also see a picture relating to
the word. In the exercises with embodied teachers (on-screen avatar
and physical robot) they would also see the
facial features of the word pronunciation (visible speech). The user would then get an
opportunity to pronounce the word themselves. In order to move on to the next word, the
word had to be pronounced correctly. When all words had been
introduced, a test would be given where the user matched the word the
computer was saying with the correct picture.

Because of the difficulty of automatically judging the correct user
pronunciation, the application was set with a human in the loop
(wizard-of-oz). The only task of the human (which was completely
hidden to the user) was to press a key to indicate good or bad
pronunciation, all other aspects of the interaction was autonomous. 

The application was built using the IrisTK framework \cite{IRISTK},
which is a Java-based framework for constructing multi-modal dialogue
systems.  IrisTK provides an API for developers including designing
flow based dialogue systems, controlling an animated face and modules
for using speech synthesizers. 

\begin{figure}

\includegraphics[width=0.5\textwidth]{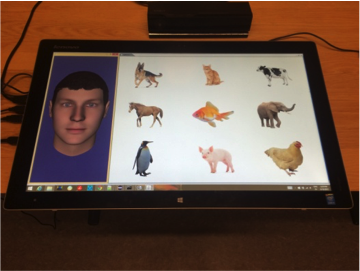}
\includegraphics[width=0.5\textwidth]{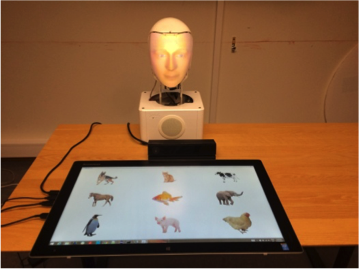}
\caption{The setup for the two interaction conditions \emph{screen} (left)
  and \emph{robot} (right). The setup for the \emph{disembodied}
  condition was identical to the one on the right, minus the robot.}
\label{fig:setup}
\end{figure}

The Robot used was Furhat \cite{FURHAT}, which is a robotic head with a
moveable neck with two degrees of freedom and an animated face that is
projected onto a plastic mask. The Furhat robot is seamlessly
integrated with the IrisTK framework which makes switching between
on-screen animated face and physical robot a matter of changing one
line of code.  Furthermore, the facial animation used in Furhat 
is the same as that used in IrisTK which makes comparisons between the
two presentation conditions straightforward. All three interaction conditions used
the same voice (CereProc voice ``William'', except from the Russian words,
which were uttered by Mac OS X voice ``Yuri''). The system interaction
took place via a 27-inch touchscreen laying flat on the table in front
of the user, see figure \ref{fig:setup}.

\section{Method}

The language learning exercise was conducted in the three different
interaction conditions: disembodied voice (disembodied), on-screen
avatar (screen) and physical robot (robot). Each subject went through
each condition in randomised order. There were three wordlists of nine
words, one wordlist being taught in each condition. The pairing of
wordlist and interaction condition was varied across subjects. 

Subjects were introduced to the different words and asked to repeat
them as described above, during a learning phase. When they had
managed to pronounce all words correctly they entered the test phase
where they were given a word (spoken) and were supposed to select the
corresponding picture on the touch screen, where the number of correctly recalled
words were counted.  

Time spent on each condition (training + test) was approximately five minutes, yielding
a total time of about 15 minutes for each subject.

Fifteen subjects participated in the study. They were all students at
KTH and none of them had any previous knowledge of the Russian
language. 

\section{Results}

\begin{figure}

\includegraphics[width=0.9\textwidth]{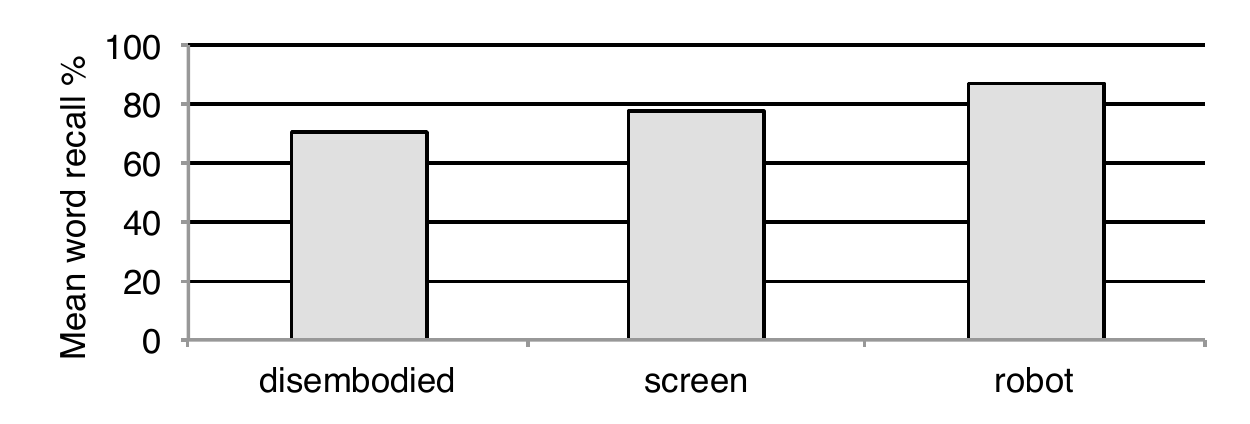}
\caption{Mean test scores as \% recalled words in the vocabulary test}
\label{fig:bars}
\end{figure}

When the mean word recall scores from three interaction conditions are compared,
the disembodied condition yielded the lowest score (42.7\%)
followed by the screen condition (46.7\%) and the robot condition
(52\%), as can be seen in figure \ref{fig:bars}. 

A \emph{Friedman test} was conducted on the test scores, which revealed a
significant effect of interaction condition ($x^2(2) = 13.43$ and $p = 0.0012$).
The Friedman test only reports wether or not there is a significant
effect of condition, and ranks the conditions in order. The mean rank
for the three conditions, where a higher rank is better, were
\emph{disembodied: 1.4}, \emph{screen: 1.9} and\emph{robot: 2.7}. 
A post hoc Wilcoxon signed-rank tests was applied on each pair
of conditions in order to find between which pairs the effect is
present (\emph{disembodied vs. screen}, \emph{disembodied vs. robot} and \emph{screen vs. robot}). Because there are multiple comparisons being made, a Bonferroni correction
must be applied to the seignificance level threshold such that $p \le
0.05 / 3 = 0.017$ for the result to be statistically significant.
The Wilcoxon signed-rank comparisons revealed a significant difference
in word recall between the conditions disembodied and robot as well as
between screen and robot:

\begin{itemize}
\item \emph{disembodied vs. screen}: $p=0.09102$ which \emph{is not}  significant at $p \le 0.17$
\item \emph{disembodied vs. robot}: $p=0.00222$ which \emph{is} significant at $p \le 0.17$
\item \emph{screen vs. robot}: $p=0.01140$ which \emph{is} significant at $p \le 0.17$
\end{itemize}

\section{Discussion}

We can see that there are significant effects on learning when the virtual tutor takes
the step from screen into the physical world. Even though the same
facial animations are used, a robotic face yields better individual
scores in relation to the other exercises. 

What could be observed, as one likely factor behind this effect,
is that more time was spent with Furhat than with the other setups. 
This can perhaps be related to what can be seen in theories around
motivation - the motivation for doing a task increases when the task
performed is enjoyable or fun. 

It can be speculated that because the experience
of a robot face is new and perhaps unique, this may affect the user in a way which could increase
the interest in the task and thus increase the task motivation, giving
a better learning experience. 

A post-trial questionnaire \cite{wedenborn4034} revealed that users
indeed found the robot version of the system to be more entertaining
and engaging than the other conditions.  In a free comment section of the questionnaire, some of the users expressed that they felt a stronger personal
connection with Furhat (i.e. the robot) and was hoping for more dialogue to be
exchanged with him. This suggests that it is not only the
news value and entertainment factor that sets the experience with the
robotic face apart, but
perhaps the robot in this case is causing users to be more emotionally invested in the experience. The extrinsic motivation could also come in to play
in such a situation, where you are more afraid of the punishment of a
bad grade - in this case a bad score on the test - coming from something you have a more human like connection with.  

\bibliographystyle{plain}
\bibliography{paper.bib}

\end{document}